# A possible origin of quantum correlations

*Alexander V. Belinsky*[*] *and Michael H. Shulman*[§]

[*]Physics Faculty, Lomonosov Moscow State University,
Vorob'evy Gory, Moscow 119991, Russia
e-mail: belinsky@inbox.ru

[§] e-mail: shulman@dol.ru

[*]Corresponding author e-mail: belinsky@inbox.ru

## Abstract

We intend to eliminate the known conflict between relativity and quantum mechanics. We believe the "instant" correlation between entangled distant quantum particles can be explained by the fact that in a *laboratory reference frame* the photon traveling duration is positive and finite while its *proper* (in vacuum) traveling duration is equal to zero. In the latter case, any two events that are separated (in a laboratory reference frame) by an arbitrary finite distance can be considered as simultaneous ones. So, the photon nonlocal correlation turns out to be a relative property and may be explained like known twins paradox in relativity. In such a situation, any standard causal interaction between the correlated particles is absent in a laboratory reference frame; however, some specific mutual couple appears between them; this couple is strictly oscillating without some oriented energy or/and information transferring. We also motivate the basic hypothesis extension on quantum particles having nonzero masses.

## 1. Introduction

The common opinion exists now that the theory of relativity interdicts any physical interaction propagating with a superluminal speed over any (arbitrary large) distances and conflicts with quantum mechanics (QM) (the known Einstein-Podolsky-Rosen paradox). This was confirmed by a number of experiments (see, e.g., [1, 2]), hence the insuperable conflict arises between QM and relativity. In 1990 John Bell pointed out [3]:

*We have the statistical predictions of quantum mechanics, and they seem to be right. The correlations seem to cry out for an explanation, and we don't have one.*

Further, he expressed hope:

*... here, I think we have a temporary confusion. It's true that it is sixty years old, but on the scale of what I hope will be human existence, that's a very small time. I think the problems and puzzles we are dealing with here will be cleared up, and we will look ba,ck on them with the same kind of superiority, our descendants will look ba,ck on us with the same kind of superiority as we now are tempted to feel when we look at people in the late nineteenth century who worried about the ether. And Michelson-Morley ..., the puzzles seemed insoluble to them. And came Einstein in nineteen five, and now every schoolboy learns it and feels ... superior to those old guys. Now, it's my feeling that all this auction at a distance and no action at a distance business will go the same way. But someone will come up with the answer, with a reasonable way of looking at these things. If we are lucky it will be to some big new development like the theory of relativity. Maybe someone will just point out that we were being rather silly, and it won't lead to a big new development.*

Below we consider several examples of physical nonlocal effects and show how the correspondent paradoxes of QM can be explained just using relativity.

## 2. The Wheeler "Galactic" Paradox

John Wheeler proposed [4] such a gedanken experiment (Fig. 1). Let a distant quasar Q emit a photon that travels to Earth during billions of years. Due to the gravitational lens action of a giant galaxy G, the photon generated from a quasar (Q) has several possible paths to reach the terrestrial telescope T through a Mach-Zehnder interferometer placed at its entry.

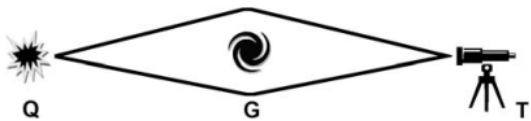

**Fig. 1.** Due to the gravitational lens action of a giant galaxy (G) the photon generated from a quasar (Q) has several possible paths to reach the terrestrial telescope (T) through a Mach-Zehnder interferometer placed at its entry.

At the entry of the telescope T, there is a Mach- Zehnder interferometer. One can insert (or not) an additional input 50% beam splitter into this interferometer. When the beam splitter is removed, the detectors allow one to determine through which path the photon propagated, but the interference pattern cannot be observed. In contrast, when the beam splitter is inserted, we cannot obtain information on the photon way[1], so the interference pattern will emerge. The paradox essence consists in the fact that the choice between the interfering and noninterfering patterns is made in the *last* time moment when the photon already finishes its billion years travel. One can consider this effect as nonlocal phenomenon - the emitted photon cannot "know" in advance if we plan to insert the beam splitter into the interferometer.

## 3. The Tetrode Paradox and Concept of Direct Particle Interaction

In the well-known paper [5], the remarkably deep thought of Hugo Tetrode[2] is quoted [6]:

*"The Sun would not radiate if it were alone in space and no other bodies could absorb its radiation ... If, for example, I observed through my telescope yesterday evening that the star which, let us say, is 100 light years away, not only did I know that the light that it allowed to reach my eye was emitted 100 years ago, but also the star or individual atoms of it knew already 100 years ago that I, who then did not even exist, would view it yesterday evening at such a such a time... "*

But how can a distant star "foreknow" where and when the emitted photon will be detected in the future? From our point of view, this is remarkable example of nonlocality that is very close to the above "galactic" Wheeler paradox.

In order to substantiate the Tetrode thesis, the version of "instant" (direct) interaction between electrons was proposed in [5] along with all possible (in future) absorbers of the emitted radiation. This idea, particularly, allows one to deduce the so-called "radiative reaction in field theory," but one has to use some complicated representation of combinations of retarded and advanced waves. As we believe, one can consider direct interaction between particles ("interaction at a distance") as an example of nonlocality.

---

[1] Inserted beam splitter plays the role of a "quantum eraser" as after photon passes through it one cannot in principle determine the preceding way.
[2] Hugo Martin Tetrode (1895 – 1931) was a Dutch theoretical physicist who contributed to statistical physics, early quantum theory, and quantum mechanics.

## 4. Duration and Length Contraction in Relativity

Note that Wheeler analyzes the above situation in a laboratory reference frame (LRF) exclusively. However, let us analyze this situation in some moving reference frame. At first, we suppose that a rocket (not a photon!) travels (with a subluminal speed) from the quasar Q to the terrestrial telescope T. While we observed this travel from Earth (in LRF), its duration was, say, one billion years. However, in the co-moving reference frame which is coupled with the rocket, the proper travel duration and length are decreased by the factor $1/\sqrt{1-v^2/c^2}$, where $v$ is the speed of the rocket and $c$ is the speed of light. The faster the rocket moves, the bigger the duration and length contract, but the event cause-and-effect relations remain the same: the rocket finish is also preceded by its start.

But what if we change the rocket or another object moving with a subluminal speed by a photon that has the *speed of light*?

As a rule, physicists do not use luminal reference frames. For example, the well-known textbook on quantum electrodynamics [7] states that the rest frame cannot exist for a particle having zero mass, because it moves with the speed of light in an arbitrary reference frame. However, it is true only in Minkowski space–time; outside of it, such a necessity appears and is really used by physicists. For example, when a particle is dropping on a black hole, its velocity really cannot exceed the speed of light c (and the dropping duration streams to infinity). However, in a transformed coordinate system [8], one can introduce the co-moving reference frame, where on the black hole event horizon the velocity of the dropping particle becomes (and then remains) bigger than the pointed out value c, and the dropping duration is finite. Note, that the new temporal and spatial coordinates should be expressed through temporal and spatial coordinates of the distant coordinate system (this is not trivial), but a possibility appears to describe processes inside the black hole.

Anyway, one can accept that the photon's case is ultimate; in such limit case (say, "from the photon point of view") the proper duration and length of the photon travel become equal to zero, and start and finish events turn out to be simultaneous and separated by zero spatial distance.

Due to this, the time moment of start of a photon emitted by the quasar and time moment of finish when it passes (or not) through a beam splitter at the telescope entry is the same time moment of its proper time. Any contradiction is absent in the statement that the photon made the choice to behave as particle or as wave just in the time moment when the beam-splitter absence or presence in the telescope forced it to make this choice. Within the framework of relativity, we are seeing that such a paradox is not only possible (photon emission and absorption events coincide in its proper time) but inevitable like the twin paradox.

We believe that we meet here practically the same paradox. On the one hand (as we noted above), how can a distant star "foreknow" where and when the emitted photon will be detected? On the other hand, we come to the remarkable idea that the photon is only a link between two (maybe, distant) atoms. We believe also that such idea does not contradict the laboratory-observer opinion according to which this photon was emitted having random orientation and during some finite time meets a random absorber.

## 5. Reviewing of Cause-and-Effect Relations

So, for a photon (not for a rocket) from its "point of view" the cause-and-effect relations (between start and finish events) cannot be adequate ones; these events become simultaneous ones — one from them cannot be a consequence of another or be preceded by it. Let now the observer be at the origin of the Minkowski coordinate system (initial four-event) and consider some other four-event.

- If this other four-event is inside of the light cone, then the four-distance between these events turns out to be real.
- If this other four-event is placed on the light cone itself, then the four-distance between these events turns out to be zero.
- If this other four-event is outside of the light cone, then the four-distance between these events turns out to be imaginary.

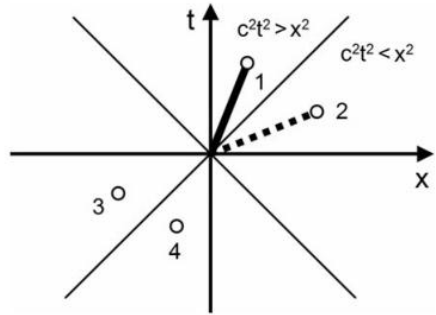

**Fig. 2.** Light cone and four-events. Events 1 and 4 can be linked with the coordinate system origin by a cause-and-effect relation, i.e., an oriented information and/or energy transferring between them is possible (nonperiodic process). Events 2 and 3 cannot be linked with the coordinate system origin by a cause-and-effect relation, i.e., an oriented information and/or energy transferring between them is impossible, as one can see. However, some mutual couple between them can support strictly periodic process. In this, we arrive at a resolution of the conflict between relativity and QM.

In [9], we pointed out that the transition from causal interaction between events inside of the light cone (timelike four-distance) to some mutual couple between events separated by light cone boundaries (space-like four-distance) can be described by a transition from nonperiodic processes to strictly periodic ones. At such a transition, any oriented (in time and space) information and/or interaction energy transferring disappears, although this energy root-mean-square turns out to be positive. This means that a two-event mutual-couple exists; however, one of them cannot be a cause nor a consequence of the other. Indeed, let a photon emitted by a lamp trigger a bomb explosion. We usually consider (in LRF) the photon emission event as the cause and the bomb explosion as the consequence, i.e., they present in LRF two individual events separated by a positive time interval. However, from the photon point of view, they are not two individual events; they present one common event, they are simultaneous, and one of them cannot be preceded by the other. Due to this, we believe it is more correct to call them coupled by a quantum correlation. It is well known [10] that such correlations may appear during transition through a light barrier (event horizon) when a particle falls on a black hole.

Thus, when we considered the above Wheeler paradox in a LFR, the photon traveling duration between start and finish time moments seemed to be positive. However, the proper travel duration is zero. Because of that, a photon start time moment (when it leaves the quasar) and finish time moment (when it passes or not through a beam splitter at the telescope entry) present the same time moment of its proper time. So, any contradiction is absent in the statement that the photon "chose" to behave as a particle or as a wave just at the same time moment when the beam-splitter presence or absence at the entry of the telescope forced it to make this choice. As relativity states, such a paradox is not only possible but necessary.

**6. "Delayed" Choice of the Photon Behavior**

In the above situation concerning the photon's behavior, the Wheeler idea of "delayed choice" was realized, according to which the decision on a measurement configuration is made at the final (not initial) stage of the photon-propagation process. Thus, one has in mind, the events chronology used in a LRF; however, this is not true from the "photon point of view."

In two experimental versions [11] (Vienna, 2007 and Canary Islands, 2008), the hybrid entangled photon-pair source emits path-polarization entangled photon pairs. The "system" photons are propagating through an interferometer on the right side, and the "environment" photons are subject to polarization measurements on the left side. The choice to acquire which-way information or to obtain interference of the system photons is made under Einstein locality (using random number generator), so that there are no causal influences (in a LRF) between the system photons and the environment photons.

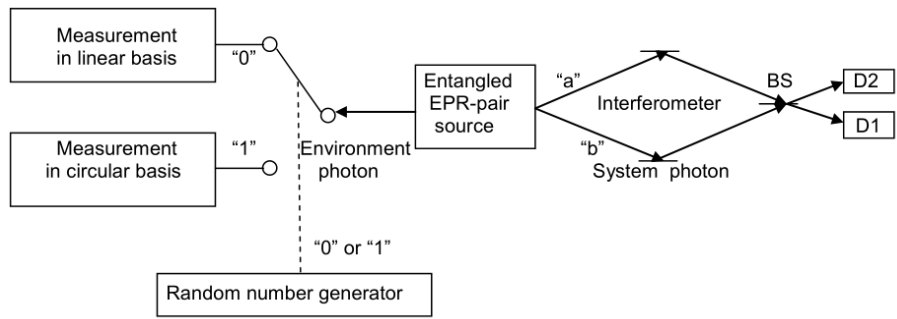

**Fig. 3.** The concept of quantum eraser under Einstein locality conditions applied in [11].

The purpose of that experiment (see Fig. 3) was to manipulate the behavior of the system photon using the measuring on the environment photon. When one measures the environment photon in linear basis, the which-way information is conserved for the system photon, and an interference pattern cannot be observed at the output of the interferometer. In contrast, when one measures the environment photon in circular basis, the which-way information turns out to be erased, and the interference pattern of the system photon appears, so, we have wavelike behavior of the system photon.

If the travel length of the system photon (at the right) to the detectors is larger than the travel length (at the left) of the environment photon to the switch, then the paradox is absent – the manipulation event occurs earlier and interference pattern appears (or not) at the interferometer output, i.e., the consequence is preceded by the cause. However, if the travel length of the system photon (at the right) to the detectors is shorter than the travel length (at the left) of the environment photon to the switch, then in a LRF a paradox seems to appear, as the author of [11] believe – the environment photon manipulation events turns out to be later than when the interference pattern appears (or not)[3].

To clear the situation, we consider the famous twin paradox in the version adapted to two photons. Let us have a source S of an (entangled) photon pair and two perfect optical fibers having two (perhaps) different lengths $L_1$ and $L_2$. The ends of the fibers are disposed close enough to one another, where a detector D of photons is placed. To be more specific, let $L_2 > L_1$.

We consider the situation in a LRF. Let the photon (entangled) pair be generated in the time moment $T_0 = 0$. The first photon will arrive at detector D in the time moment $T_1 = L_1/c$, and the second one will arrive at detector D in the time moment $T_2 = L_2/c$, where c is the speed of light. The difference in time moments is

$$\Delta T = (L_2 - L_1)/c$$

Note that for *each* photon the proper travel duration will be zero, and the difference in time will also be zero, $\Delta T' = 0$. In other words, if both photons had watches synchronized at the pair-emission time moment, the difference between arrival time moments was also zero.

Because of that, we believe that the paradox may be explained like the preceding case. All the considerations of the authors of the experiment (like the EPR experiments) are based on the analysis in a LRF; however, if one considers the events using the photon *proper* time, than the travel duration (and spatial length) between the measurements on the system photons and environment photon will go to zero, and hence any mismatch between them is impossible.

---

[3] The authors of [11] refer to the experimental results they reached.

## 7. The Hypothesis Extension on Quantum Particles Having Nonzero Masses

Although the nonlocality phenomenon seems to be actually relative depending on the reference frame, it might not be true for a massive particle which cannot propagate with speed of light. Note that in many experiments with delayed choice for massive particles (not for photons), only photons turn out to be the main participants (see, e.g., [12]). However, for EPR experiments with entangled electrons moving with subluminal speed, such explanation seems at first sight to be not acceptable.

When one considers two flying-off entangled massive particles, it is impossible to measure precisely the position of the first particle and the momentum of the second one due to the Heisenberg uncertainty relation in a LRF. Such argumentation is true if one can think of the particles as perfect mechanical balls. However, we will try to demonstrate that the entangled particles are not perfect mechanical balls without any links.

In fact, massive quantum particles (particularly electrons) have also wave properties, not only corpuscular ones. For example, Dirac in 1928 presented [13] a relativistic description of the electron wave function through a system of four differential equations for four spinors, where one pair corresponds to positive energy and the second pair corresponds to negative energy. In each pair, one of the spinors corresponds to a direction, and the other spinor corresponds to the opposite direction. Furthermore, the operators of electron speed components do not commute, and the eigenvalues of each of them at a measurement have moduli exactly equal to the speed of light. Schrödinger in 1930 explained [14] such a paradox by the existence of two electron speed components – "usual" (slow) and "quickly oscillating" with the frequency corresponding to the electron de Broglie wave period. He wrote also that the square of each speed component can only take the value $c^2$ and also be the average (mathematical expectation) over many measurements on the same wave pocket. The speed component itself can take the values $\pm c$. However, its expectation may be (and generally is) smaller. Meanwhile, one wonders how the charged-cloud center of gravity may move with a subluminal velocity. It is possible only because it does not move uniformly.

Similar representations of a real electron that is composed of two massless components "zig" and "zag" are described by Penrose [15]:

*...Dirac spinor, ...with its four complex components, can be represented, as a pair of two-spinors...The Dirac equation can then be written as an equation coupling these two two-spinors, each acting as a kind of 'source' for the other, with a 'coupling constant' $M/\sqrt{2}$ [M is the mass] describing the strength of the 'interaction' between them ...From the form of these equations, we see that the Dirac electron can be thought of as being composed of two ingredients .... It is possible to obtain a kind of physical interpretation of these ingredients. We form a picture in which there are two 'particles,' ...each of which is massless and where each one is continually converting itself into the other one ...Being massless, each of these should be traveling with the speed of light, but we can think of them, rather, as 'jiggling' backwards and forwards where the forward motion of the zig is continually being converted to the backward motion of the zag, and vice versa. In fact, this is a realization of the phenomenon referred to as 'zitterbewegung', according to which the electron's instantaneous motion is always measured to be the speed of light, owing to the electron's jiggling motion, even though the overall averaged motion of the electron is less than the light speed. Each ingredient has a spin about its direction of motion, of magnitude ?/2, where the spin is left-handed in the case of the zig and right-handed for the zag ...In this interpretation, the zig particle acts as the source for the zag particle and the zag particle as the source for the zig particle, the coupling strength being determined by M. In the*

*total process, we find that the average rate at which this happens is (reciprocally) related to the mass coupling parameter M; in fact, this rate is essentially the de Broglie frequency of the electron.*

Finally, in [16] it is even more clearly formulated that such "zitterbewegung" corresponds to the stationary state of an electron as a superposition of two eigenstates of a speed operator having eigenvalues ±c. As a result, the "effective" speed of electron is

$$v_z = c^2 \frac{p_z}{E} + \frac{i\hbar c \dot{v}_{z0}}{2E} \exp(-2iEt/\hbar) = (v_z)_{mean} + (v_z)_{osc}$$

where $p_z$ is the momentum projection, $E$ is the particle energy, and $\dot{v}_{z0}$ is the value of $\dot{v}_z$ at t = 0. Just the mean speed ( $(v_z)_{mean}$ is determined by the *actually measured* value of the particle momentum $p_z$. The mean speed and momentum directions are the same only in stationary states with positive energy and are opposite one to another when the energy is negative.

So, when we consider the entangled electrons spins, we meet necessarily a nontrivial wavelike (oscillating) process where the couple between components (not linked with the real motion of electron) is specified by the speed of light. Hence, two entangled particle, as we noted above, cannot be represented as two independent mechanical balls but should be represented as components of a nonlocal wave domain (Fig. 4) spaced (in LRF) between initial and final points. However, from the conditional photon point of view, this domain simply "contracts" to a single point.

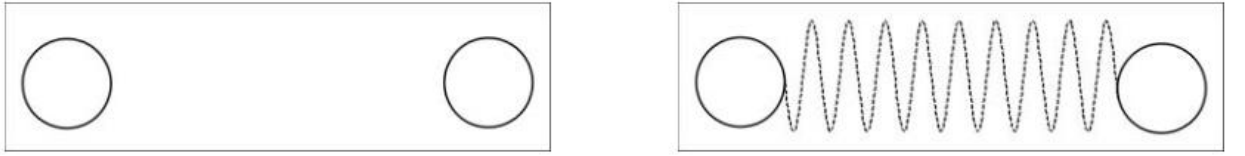

**Fig. 4.** Classical (perfect mechanical balls) and quantum (real quantum particles) representations of a pair of (entangled) particles.

In other words, in LRF we have some "*effective*" particle motion with *subluminal mean velocity*, but an instant mutual coupling between boundaries appears from the electron point of view. In practice, such duality corresponds to the de Broglie–Bohm pilot wave concept [17,18] and gives immediately the possibility to describe reality *quantitatively* like quantum-mechanics laws.

**8. De Broglie–Bohm Interpretation**

In 1923 L. De Broglie published [17] the hypothesis that massive particles (like electrons) have also wave properties (not only corpuscular) as well as photons. Thus, such particle motion is guided by some pilot wave. In 1952 D. Bohm developed his idea and published two coupled papers [18]. Bohm proposed there to change the *single* **Schrödinger** equation for *complex* wave function $\Psi = \sqrt{\rho} \exp(iS/\hbar)$ with an external potential $V$

$$i\hbar \frac{\partial \Psi}{\partial t} = \left(-\frac{\hbar^2}{2m} + \nabla^2 + V\right)\Psi$$

by a system of two coupled equations for two *real* quantities – the square of the amplitude modulo $\rho$ and phase $S$ of the wave function:

$$\frac{\partial \rho}{\partial t} + \nabla \cdot \left( \rho \frac{\nabla S}{m} \right) = 0, \qquad \frac{\partial S}{\partial t} + \frac{(\nabla S)^2}{2m} + V + Q = 0.$$

Here

$$Q \equiv \frac{\hbar^2}{4m} \left[ \frac{1}{2} \left( \frac{\nabla \rho}{\rho} \right)^2 - \frac{\nabla^2 \rho}{\rho} \right]$$

is so-called *quantum potential*. The first equation (without $Q$) is the usual equation of continuity for the probability density $\rho$, while the second one (including $Q$) describes the phase evolution guiding *quantum*-particles motion.

The phase in the Bohm equation turns out to be a nonlocal parameter; its dynamics depends on specific quantum potential that itself depends exclusively on non-uniformity density distribution in space. The quantum potential presence differentiates the quantum description from the classical one, which does not contain this quantity in any analogy. Generally, the quantum potential entangles the particles between them; hence it corresponds to the fact that the particle individual trajectories (that have a physical sense in the Bohm's interpretation) are not independent of one another and cannot be described by individual independent wave functions.

Bohm believed [18] that the pilot wave phase should be considered as a "hidden parameter" in the von Neumann sense [19]. Von Neumann pointed out that such parameters as hypothetically precise quantum statistical descriptions cannot exist. In turn, in 1964 J. Bell showed [20] that the existence of hidden parameters leads to several inequalities [21] that are violated by quantum mechanics, in full correspondence with experiments.

Bell analyzed the possible origin of the inequality violation and assumed that it was due to quantum interaction nonlocality. Because in the Bohm theory the wave-function phase is nonlocal (see [22]), from Bell's viewpoint this theory does not contradict the von Neumann statement on the impossibility of "hidden parameters" in quantum mechanics.

Note that the proof itself of Bell's results was critically discussed. For example, in [23] it is written

"... *that violation of Bell's inequality might be interpreted not only as an evidence of the alternative – either nonlocality or "death of reality" (under the assumption that the quantum mechanics is incomplete).*

*Violation of Bell's type inequalities is a well-known sufficient condition of incompatibility of random variables – impossibility to realize them on a single probability space.*

*Thus, in fact, we should take into account an additional interpretation of violation of Bell's inequality – a few pairs of random variables (two-dimensional vector variables) involved in the EPR–Bohm experiment are incompatible. They could not be realized on a single Kolmogorov probability space. Thus, one can choose between: a) completeness of quantum mechanics; b) nonlocality; c) " death of reality"; d) non-Kolmogorovness. In any event, violation of Bell's inequality has a variety of possible interpretations.*"

Anyhow, Leggett [24] showed later that nonlocal theories with hidden parameters of several types (nonlocal couple of distant measurers) also are limited by inequalities, which may be violated by quantum-mechanical predictions.

We believe that the Broglie–Bohm-concept correctness is just the theoretical consequence of our model where the nonlocality problem is simply absent "from the viewpoint" of any component moving with the speed of light, and all the four-events contract to the common four-point.

The researchers working within the frame of the de Broglie–Bohm approach reached a number of deep, important and general results. They calculated the diffraction and interference patterns for a set of standard cases as well as for nontrivial ones.

For example, in [25], the numeric simulation results for two-slit experiments with electrons are presented. In [25], the evolution of the probability density from the source to the detection screen is deduced; the calculations were made using the method of Feynman path integrals. The wave function after the slits was deduced from the values of the wave function at slits A and B. Also, the authors reached the analytic solution for the wave function in the Stern–Gerlach experiment where they calculated the decoherence time and the diagonalization of the density matrix. This solution requires the calculation of the Pauli spinor with a spatial extension as the equation

$$\Psi(z) = (2\pi\sigma_0^2)^{-1/2} \exp(-\frac{z^2}{4\sigma_0^2}) \begin{pmatrix} \cos\frac{\theta}{2} e^{-i\varphi/2} \\ \sin\frac{\theta}{2} e^{-i\varphi/2} \end{pmatrix}$$

while quantum mechanics textbooks do not take into account the spatial extension of the spinor and use the simplified spinor

$$\Psi(z) = \begin{pmatrix} \cos\frac{\theta}{2} e^{-i\varphi/2} \\ \sin\frac{\theta}{2} e^{-i\varphi/2} \end{pmatrix}$$

In [25], it was shown that different evolutions of the spatial extension between the two spinor components play a key role in the explanation of the measurement process and allow one revisit the Stern–Gerlach experiment.

Additionally, in [25] the Bohm's version of the Einstein–Podolsky–Rosen experiment was investigated. The following causal interpretation of EPRB-experiment is proposed: As the authors assume, at the creation of the two entangled particles A and B, each of two entangled particles has an initial wave function with opposite spins. Then the Pauli principle tells us that the two-body wave function must be antisymmetric. Thus, one can consider that the singlet wave function is the wave function of a family of two fermions A and B with opposite spins, where the direction of the initial spins A and B exists but is not *known*. This is not the interpretation followed by the Bohm school in the interpretation of the singlet wave function; they do not assume the existence of different functions for each particle, hence they assume a zero spin for each particle at the initial time and that the spin modulus varied during the experiment from 0 to ħ/2. In contrast, the authors of [25] assume that at the initial time the spin of each particle (given by each initial wave function) and the initial position of each particle are known.

The de Broglie–Bohm concept may be applied also to the physics of electromagnetic radiation and other wave processes (e.g., acoustics), not only to the particle physics. For example, in [26], the light pulse in a waveguide within the small angle or the paraxial approximation was considered. Assuming that the optical axis is oriented along the z-axis and the electromagnetic field passing through the waveguide is time harmonic, the field can be approximated by a plane wave along the z-direction modulated by a certain complex-valued amplitude,

$$\Psi(r) = \psi(r)\exp(ik_z z)$$

where $k_z = n_0 k$, $k = 2\pi/\lambda$, $\lambda$ is the light wavelength in vacuum, and $r$ is the bulk refractive index. Substituting this expression into the Helmholtz equation, one arrives at an equation isomorphic to the Shrödinger equation, with $z$ playing the role of the evolution parameter (rather than the time t). The last equation has been used, for example, to study the design of waveguides with optimum conditions of light transmission.

Note that these calculation results are fully in coincidence with standard quantum-mechanics predictions.

## 9. Conclusions

In this paper, we departed from a simple statement. When some experiment is considered in different reference frames where time currency is sufficiently different, one has to take into account the comparison of the results should have an objective character for the same four-events. However, in this case, several paradoxical situations may appear, and several properties (particularly, nonlocality) may turn out to be relative ones. For example, in the twin paradox, the ages of the earth and astronaut are compared at the same four points of space–time (initial and final). However, their age increments are calculated in different reference frames, exactly as relativity requires; so, the paradox should appear and really does!

From our point of view, we observe a similar situation in a quantum experiment with nonlocal correlations between photons. The results established in different reference frames that should be compared at the same initial and final conditions lead to apparent paradoxes. However, such paradoxes are inevitable and correspond to the object properties in different reference frames. Probably, such couple instantaneity (from the photon viewpoint) may also explain the role of the which-way information, because this information appears just at the same time moment at which the result is detected.

Concerning an entangled pair of quantum particles having nonzero mass and propagating with subluminal velocity, we already considered the arguments in favor of this hypothesis in the previous section. The same arguments should be applicable for teleportation experiments.